\documentclass[traditabstract]{aa} 
\usepackage{graphicx}
\usepackage{epsfig,amsmath,amssymb}

\usepackage{txfonts}
\def\msol{M_\odot}
\def\mjup{M_{\rm J}}
\def\msolyr{M_\odot \rm {yr}^{-1}}
\def\msun{M_\odot}

\def\mstar{M_\ast}
\def\lstar{L_\ast}
\def\minit{M_{\rm init}}
\def\nburst{N_{\rm burst}}
\def\tburst{\Delta t_{\rm burst}}
\def\tquiet{\Delta t_{\rm quiet}}
\def\mdotburst{\dot{M}_{\rm burst}}
\def\mdotquiet{\dot{M}_{\rm quiet}}
\def\rstar{R_\ast}

\def\te{T_{\mathrm{eff}}}
\def\tc{T_{\mathrm{c}}}

\def\mdot{\dot{M}}
\def\tdot{\tau_{\rm \dot M}}
\def\tkh{\tau_{\rm KH}}

\def\be{\begin{equation}}

\def\ee{\end{equation}}
\def\mjup{M_\mathrm{{Jup}}}

\def\simgr{\,\hbox{\hbox{$ > $}\kern -0.8em \lower 1.0ex\hbox{$\sim$}}\,}
\def\simle{\,\hbox{\hbox{$ < $}\kern -0.8em \lower 1.0ex\hbox{$\sim$}}\,}

\begin{document}
   \title{Effect of episodic accretion on the structure and the lithium depletion of low-mass stars and 
   planet-hosting stars}

%   \subtitle{}

   \author{I. Baraffe\inst{1,2} \and G. Chabrier \inst{2,1}
           %\and J. Gallardo\inst{3}
          }

 % \offprints{I. Baraffe}

   \institute{School of Physics, University of Exeter Stocker, Road, Exeter, UK EX4 4QL
   \and
   \'Ecole Normale Sup\'erieure, Lyon, CRAL (UMR CNRS 5574), Universit\'e de Lyon, France\\
             \email{ibaraffe,chabrier@ens-lyon.fr}
%\and
%Observatorio Astron\'omico Cerro Cal\'an, Departamento de Astronom\'{\i}a, Universidad de Chile, Santiago, Chile\\\email{gallardo@das.uchile.cl}
%             \thanks{}
             }

   \date{Received ; accepted}

% \abstract{}{}{}{}{} 
% 5 {} token are mandatory
 
  \abstract
  % context heading (optional)
  {Following up our recent analysis devoted to the impact of non steady accretion on the location of young low-mass stars or brown dwarfs in the Herzsprung-Russell diagram, we perform a detailed analysis
  devoted to the effect of burst accretion on the internal  structure of low-mass and solar type stars. We find that 
  episodic accretion can produce objects with significantly higher central temperatures than the ones
of the non accreting counterparts of same mass and age. As a consequence, lithium depletion can be severely enhanced in these objects.
This provides a natural explanation for the unexpected level of lithium depletion observed in young objects for the inferred age of their parent cluster. These results 
 confirm the limited reliability of lithium abundance as a criterion for assessing or rejecting cluster membership. 
 They also show that lithium is not a reliable age indicator, because its fate strongly depends on the past accretion history of the star.
	 Under the assumption that giant planets primarily form in massive disks prone to gravitational instability and thus to accretion burst episodes, the same analysis also explains the higher Li depletion observed in planet hosting stars. 
 At last, we show that, depending on the burst rate and intensity, accretion
 outbursts  can produce solar mass stars with lower convective envelope masses, at ages less than a few tens of Myr, than predicted by standard (non or slowly accreting) pre-main sequence models. 
 This result has interesting, although speculative, implications  for the recently discovered depletion of refractory elements in the Sun. 
 %Our results suggest the existence of a corelation between  the underabundance of refractory elements and the level of lithium depletion in solar type stars, i.e stars showing depletion of refractory elements should also show trends of higher Li depletion.
 }

    \keywords{Stars: formation --- Stars: low-mass --- Stars: abundances --- Accretion, accretion disks  
                    }

  \titlerunning{Effect of episodic accretion on internal structure and lithium depletion  }
  \authorrunning{Baraffe \& Chabrier}
   
    \maketitle

%
%________________________________________________________________

\section{\label{introduction} Introduction}

There is a growing consensus in the star formation community that non steady (episodic) accretion plays a dominant role during the formation of low-mass stars (see e.g Enoch et al. 2009; Vorobyov 2009; Zhu et al. 2010a and references therein). In a recent paper (Baraffe, Chabrier, Gallardo 2009, hereafter BCG09), we have suggested that episodic accretion provides a viable explanation for the observed luminosity spread in young cluster Herzsprung-Russell diagrams (HRD). The present follow-up analysis explores in more details the effects of episodic accretion on the internal structure of young low mass stars ($\leq 1\,\msol$), commonly used to derive ages of star forming regions and young clusters. We show that, depending on the accretion history, the internal structure of these objects can be strongly affected for up to a few tens of Myr (\S \ref{section_structure}). Lithium depletion and, for the partly convective stars, the size of
the convective envelope, in particular can strongly differ from the standard (non accreting) pre-main sequence model predictions (\S \ref{section_lithium}). 
In section \ref{discussion}, we examine the impact of episodic accretion on the observational signatures and show that taking  this
 process into account in the young low-mass object evolution provides a consistent explanation for the puzzling observations of strong lithium depletion in
several low-mass stars (LMS) belonging to young clusters (e.g Kenyon et al. 2005; Sacco et al. 2007) and in planet-hosting stars (Israelian et al. 2009), as well as to the recently determined peculiar abundances of refractory elements 
in the Sun (Melendez et al. 2009; Ramirez et al. 2009). 

\section{\label{section_structure} Effect of mass accretion on  the internal structure}

\subsection{Evolutionary models with accretion}

We adopt the same input physics and the same treatment of accretion as outlined in BCG09. 
%In particular, we assume that accretion onto the central object rapidly proceeds non-spherically, affecting only a small fraction of the contracting object's surface. 
%Additional details regarding the treatment of accretion process in our stellar evolution code are provided below.
In standard stellar evolution calculations, energy conservation 
equation  for a non accreting object reads as
\begin{equation}
\left( {\partial L \over \partial m}\right)_t = - T\left ({\partial S \over \partial t}\right)_m + 
\epsilon_{\rm nuc},
\end{equation}
where $m$ is the mass enclosed in a sphere of radius $r$ within the object, $S$ the specific entropy and $\epsilon_{\rm nuc}$  the local nuclear energy generation rate. For very young objects, 
only deuterium fusion 
provides a contribution to $\epsilon_{\rm nuc}$\footnote{The nuclear energy generation produced by  lithium fusion is completely negligible.}. For accretion proceeding
at a rate $\dot M$, time derivatives at fixed mass 
shell
must account for the variation of mass with time and of entropy with accreted mass, i.e. (Sugimoto \& Nomoto 1975):
\begin{equation}
\left ({\partial S \over \partial t}\right)_m = \left ({\partial S \over \partial t}\right)_q - \dot m
\left ({\partial S \over \partial m}\right)_t,
\end{equation}
with  $q \equiv m/\mstar$.
The mass ($\dot M \cdot \Delta t$)  accreted
during a timestep $\Delta t$ is assumed to be 
redistributed over the entire structure, with the 
new mass in a shell of fixed $q$ given by
\begin{equation}
m(t+\Delta t) = q \times (\mstar(t) \, + \, \dot M \,\Delta t).
\end{equation}

In the present calculations, we assume instantaneous and uniform redistribution 
of the extra source of internal energy brought by the accreted material.
In reality, mass and heat redistributions inside the accreting object
depend on the thermal properties of the accreting material. Proto low-mass stars below about 2 $\msol$ are expected to be entirely convective (Stahler \& Palla 2005). For completely 
convective objects, given the short typical convective timescales, our assumption of uniform
mass and heat redistribution should be valid, providing the entropy of accreted matter is comparable to or less than the accreting object's internal one, so that the infalling material can rapidly thermalize
with its surroundings (see discussion in Siess \& Forestini 1996; Hartmann et al. 1997). 
%This latter constraint is likely to be valid if accretion proceeds through a disk, so that a substantial fraction of the accreting energy is radiated away in the disk (BCG09).
If the aforementioned condition is not fulfilled, the accreting matter may not be able to penetrate 
very deep inside the object, possibly leading to Rayleigh-Taylor like instabilities in stably radiative regions. Siess et al. (1997) explored the effect of mass and heat redistribution by using a penetration 
function. Such an approach, however, remains highly phenomenological and thus of limited reliability.
Given the complexity of the problem and the present exploratory nature of the effect of episodic accretion on young low-mass objects, not
mentioning uncertainties in the accretion processes, we elected to stick to the simplest and probably most reasonable in most of the presently explored situations treatment, and assumed homogeneous, rapid heat and matter redistribution within the accreting body.

Omitting the fraction $\epsilon(1-\alpha)\, { G M \mdot \over R}$ of
the accretion shock energy radiated away (see Eqs. (1) and (2) of BCG09 for the definition of $\alpha$ and $\epsilon$), which does not affect
the accreting object's structure and evolution, the intrinsic luminosity of the protostar or brown dwarf is given by

\begin{equation}
\lstar =  \alpha \epsilon { G M \mdot \over R}
+ \int_M \epsilon_{\rm nuc} \, dm \,
-  \int_M T \Big\{ \left ({\partial S \over \partial t}\right)_q - \dot m
\left ({\partial S \over \partial m}\right)_t \Big\} \, dm. 
\label{eql}
\end{equation}

\noindent As mentioned above, we assume in the present calculations that the accreted matter reaches the stellar surface with a lower specific entropy than the object's internal one, which implies
$\alpha \rightarrow 0$ in Eq. (\ref{eql}). 
%This should be justified if accretion occurs essentially through a disk. 
As shown in BCG09, calculations based on this assumption provide a consistent explanation of the luminosity spread in young cluster HRDs, while high values of $\alpha$ would predict overluminous objects during a short period of time (because of the short corresponding Kelvin-Helmholtz  timescale, see discussion in BCG09). 
Addition of fresh infalling deuterium and lithium is  accounted for during the accretion process, with
interstellar mass fractions $[D]_0= 2\times 10^{-5}$ and $[Li]_0= 10^{-9}$, respectively. Depending on the accretion rate, these elements can
start to be consumed at early stages in the proto-object interior during the burst accretion phases. The implementation of the accretion process in the stellar evolution code has been validated (Gallardo 2007; Gallardo et al. 2009) by comparing our numerical results to those derived from the polytropic approach of
Stahler (1988) and Hartmann et al. (1997)  in the limit of validity of the polytropic model, {\it i.e.} 
for initial masses $\simgr$ 0.1$\msol$. 

\subsection{\label{episodic} Effect of episodic accretion on the internal structure}

As in BCG09, our accretion histories are based on the calculations of Vorobyov \& Basu (2005), obtained from gravitational instability in the accreting centrifugally supported disk. 
Starting from arbitrary initial masses $\minit$, we vary the number of bursts $\nburst$, with accretion rate $\mdotburst$ and duration $\tburst$, separated by quiescent phases of duration $\tquiet$ and accretion rate $\mdotquiet$.
We explored a wide range of parameters, inferred from the results of Vorobyov \& Basu (2005), with the aim of obtaining final LMSs within the mass range 0.1-1 $\msol$, in
order to focus on the 
typical population used to derive the ages of star-forming regions and young clusters. Brown dwarf cases have been shown in BCG09.
Our initial masses $\minit$ range from 1 $\mjup$ to 0.2 $\msol$. The typical burst number,
accretion rates, and duration lie in the ranges $\nburst$ = 2 - 30, $\mdotburst$ = 10$^{-4}$ - 5 10$^{-4}$ $\msolyr$, and $\tburst$ = 100 - 500 yr, respectively, while quiescent phases last between $\tquiet$ = 10$^3$ and 10$^4$ yr with accretion rates $\mdot < 10^{-6} \msolyr$. As mentioned in BCG09, accretion rates during the quiescent phase below this
value have no significant impact on the final internal structure of the accreting object. For the sake of simplicity, we have thus adopted $\mdotquiet$=0 in the present analysis, however, this accretion luminosity must be taken into account when comparing the final luminosity with observations of proto-objects during the embedded phase. 
The next two sections illustrate the typical results for two examples at each end of the presently studied mass range.

\subsubsection{\label{m01} Sequences leading to the formation of a 0.1 $\msol$ star}

\begin{figure}
\psfig{file=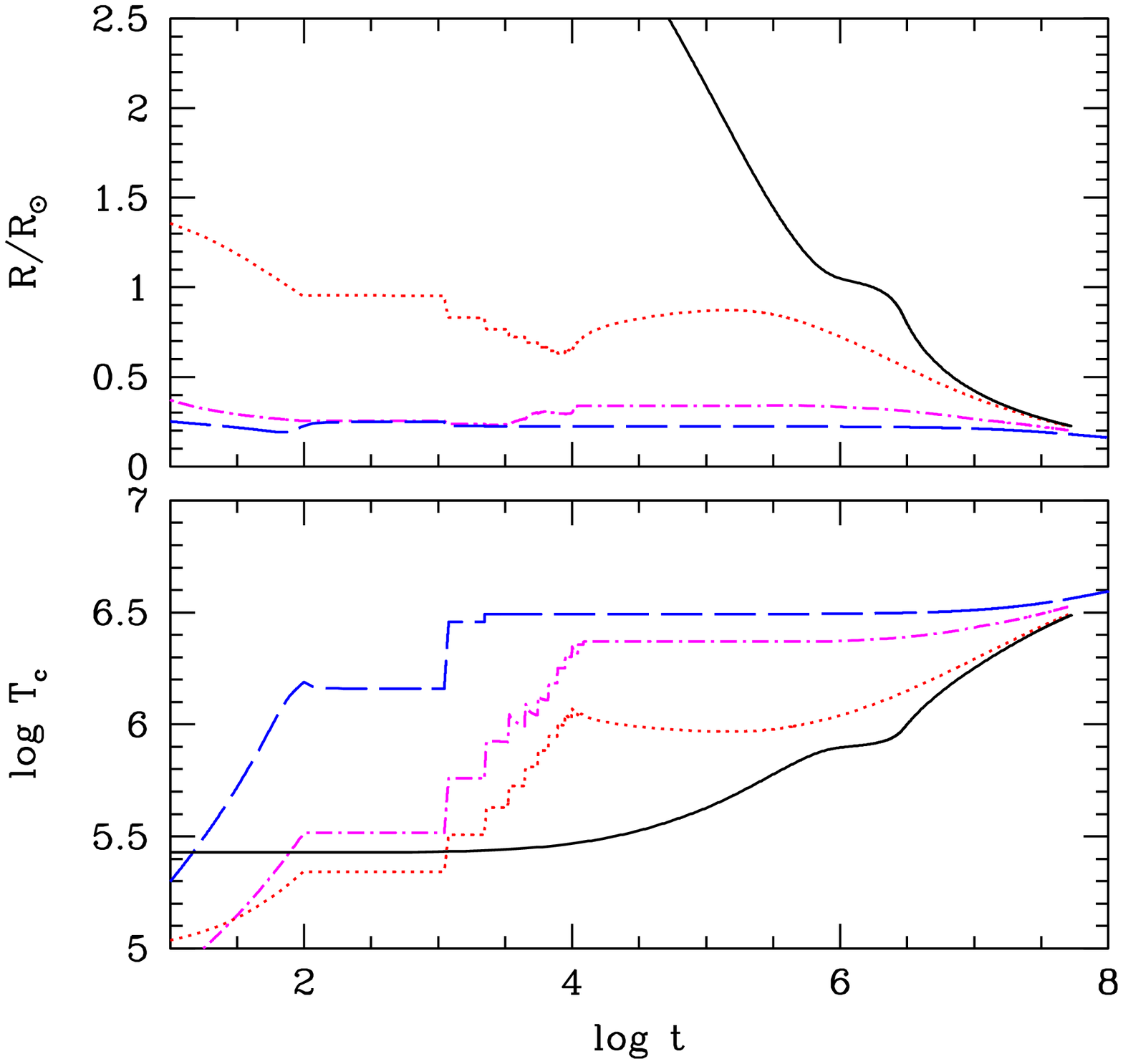,height=120mm,width=88mm}
\caption{Evolution of the radius (upper panel) and of the central temperature (lower panel) as a function of time (in yr) for models with episodic accretion and reaching a final mass of 0.1 $\msun$ (see \S \ref{m01}). Long-dash (blue): $\minit$ = 1 $\mjup$, $\mdotburst = 5 \, 10^{-4} \msolyr$, $\nburst =$2; dash-dot (magenta): $\minit$ = 1 $\mjup$, $\mdotburst = 10^{-4} \msolyr$, $\nburst =$10; dot (red): $\minit$ = 10 $\mjup$, $\mdotburst = 10^{-4} \msolyr$, $\nburst =$9. All calculations are done with $\tburst$=100 yr and $\tquiet$=1000 yr. The solid line (black) corresponds to the evolution of a non accreting 0.1 $\msol$ star.
}
\label{mrtc_mf01}
\end{figure}

Various evolutionary sequences of episodic accretion, with different initial masses, leading to a final 0.1 $\msol$ LMS are portrayed in Fig. \ref{mrtc_mf01}. 
As mentioned in BCG09, high accretion rates yield significantly more compact structures, i.e., smaller radii compared to those of non accreting objects of same mass and age. This contraction comes from the increase in gravitational energy as mass is added, yielding higher central pressures and temperatures compared with the non accreting case. 
If the  accretion timescale, $\tdot=\mstar/\dot M$, is higher than  the Kelvin-Helmholtz timescale, $\tkh=G\mstar^2/(\rstar \lstar)$, the object can relax to the radius 
it would have in the absence of accretion. In contrast, if $\tdot \le \tkh$, which
is the case for $\mdotburst =$ 10$^{-4}$ - 5 10$^{-4}$ $\msolyr$,   the structure has no time to
adjust to the incoming mass and energy, and the radius remains smaller than the non accreting
counterpart of same mass and age, as illustrated in Fig. \ref{mrtc_mf01}. The more compact structure
of the accreting object yields higher central temperature compared to the non accreting counterpart.
Figure \ref{mrtc_mf01} highlights the main effects of episodic accretion on proto low-mass objects, which can be summarized as follows.
\begin{itemize}
\item{} For a given initial mass $\minit$, the higher $\mdotburst$  during the burst phases, the smaller the radius of the object at the end of the accretion phase, hence the higher its central temperature. This can be seen by comparing the long-dashed blue curve ($\minit$ = 1 $\mjup$, $\mdotburst = 5 \, 10^{-4} \msolyr$) and the dash-dotted magenta curve ($\minit$ = 1 $\mjup$, $\mdotburst = 10^{-4} \msolyr$). 
 \item{} For a given burst accretion rate $\mdotburst$, the lower the initial mass $\minit$,
 the greater the impact on the internal structure, {\it i.e}, the smaller the radius and the higher the central temperature at the end of the accretion history. This is illustrated by the comparison between the dash-dotted magenta curve ($\minit$ = 1 $\mjup$, $\mdotburst = 10^{-4} \msolyr$) with the dotted red curve ($\minit$ = 10 $\mjup$, $\mdotburst = 10^{-4} \msolyr$).
\end{itemize} 

It is worth stressing that the sequences starting with $\minit$=1 $\mjup$ in Fig. \ref{mrtc_mf01} reach high enough central temperatures during the burst accretion episodes to completely burn deuterium, including the extra accreted fresh one. This extra supply of nuclear energy is not sufficient to balance the increase in gravitational
energy, hence the contraction of the object induced by the added matter. 
Deuterium burning starts when the central temperature $T_{\rm c}$ reaches $\sim 10^6$K. For the two sequences starting with $\minit$=1 $\mjup$ and $\mdotburst = 10^{-4} \msolyr$ and $\mdotburst = 5 \, 10^{-4} \msolyr$, this corresponds to central densities of $\rho_{\rm c}=$ 20 and 30 g cm$^{-3}$, respectively. 
 %and a D lifetime against proton capture of $\sim 100$ yr at such temperature and densities. 
 This stage is reached when both sequences reach a mass of $\sim 0.03 \msol$. 
For the sequence starting with $\minit$=10 $\mjup$ and $\mdotburst = 10^{-4} \msolyr$, the same central temperature is reached near the end of the accretion phase, for a central density of $\sim$ 2 
 g cm$^{-3}$ and a total mass  $\sim 0.08 \msol$. For this  sequence, a significant fraction of deuterium
($\sim$ 65\% of the initial mass fraction in the fully convective interior)  is still present at the end of the accretion phase, generating an important source of
nuclear energy which produces the increase in R and then decrease in $\tc$ between 10$^4$ yr and 10$^5$ yr, as shown in Fig. \ref{mrtc_mf01}. 
 
Finally, Fig. \ref{mrtc_mf01} clearly illustrates, for the sequences starting with 1 $\mjup$, the strong differences in radius and central temperature after a few Myr, up to 30 Myr, between a 0.1 $\msol$ star
produced by episodic accretion with $\mdotburst >  10^{-4} \msolyr$ and the non accreting counterpart. We also stress that, for initial masses $\minit \simgr \, 0.05 \msol$  and burst accretion rates $\mdotburst < 5 \times 10^{-4} \msolyr$, accretion history hardly affects the internal structure and location in the HRD after
$\sim$ 1 Myr, producing objects of 0.1 $\msol$ with properties {\it similar} to those predicted by standard (non accreting) pre-main sequence models  at ages $\ge$ 1 Myr.  

\subsubsection{\label{m1} Sequences leading to the formation of a 1 $\msol$ star}

\begin{figure}
\psfig{file=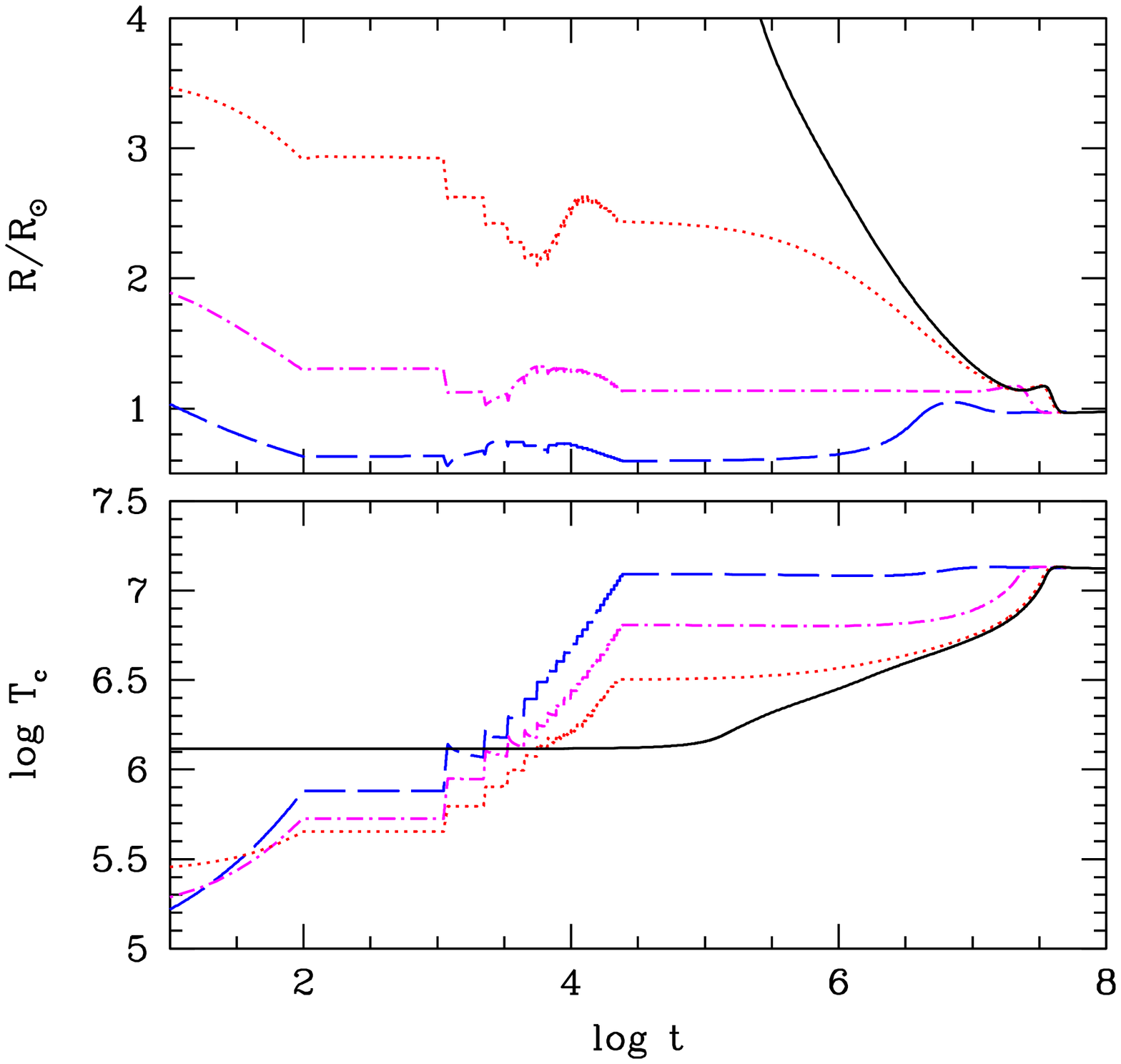,height=120mm,width=88mm}
\caption{Same as Fig. \ref{mrtc_mf01} for models reaching a final mass 1 $\msun$ (see \S \ref{m1}). Long-dash (blue): $\minit$ = 10 $\mjup$, $\nburst =$20; dash-dot (magenta): $\minit$ = 30 $\mjup$, $\nburst =$20; dot (red): $\minit$ = 0.1 $\msun$, $\nburst =$18. All calculations are done with $\mdotburst = 5 \, 10^{-4} \msolyr$, $\tburst$=100 yr and $\tquiet$=1000 yr. The solid line (black) corresponds to the evolution of a non accreting 1 $\msol$ star.
}
\label{mrtc_mf1}
\end{figure}

Figure \ref{mrtc_mf1} displays the same analysis
for episodic accretion sequences producing 1 $\msol$ objects. The figure portrays the
results with $\mdotburst = 5 \times 10^{-4} \msolyr$, which yields the strongest and most interesting
effects on the structure and on lithium depletion (see \S \ref{section_lithium}). As found previously,
the lower the initial mass, the greater the effect on the structure for a given accretion rate. For initial
masses $\minit < 0.1 \msol$ and such high burst accretion rates, the figure illustrates the strong departure in radius and central temperature of these models from the non accreting sequence up to 20-30 Myr. 
For higher initial masses and $\mdotburst < 5 \times 10^{-4} \msolyr$, the accretion history
is found not to significantly alter the properties of the newly formed 1 $\msol$ star, producing an object
with a position in the HRD and internal structure {\it similar} to the predictions of 
a non accreting evolutionary sequence at ages $\ge$ 1 Myr.

\section{\label{section_lithium} Effect on the size of the convective envelope and on lithium depletion}

\subsection{The case of fully convective stars}

\begin{figure}
\psfig{file=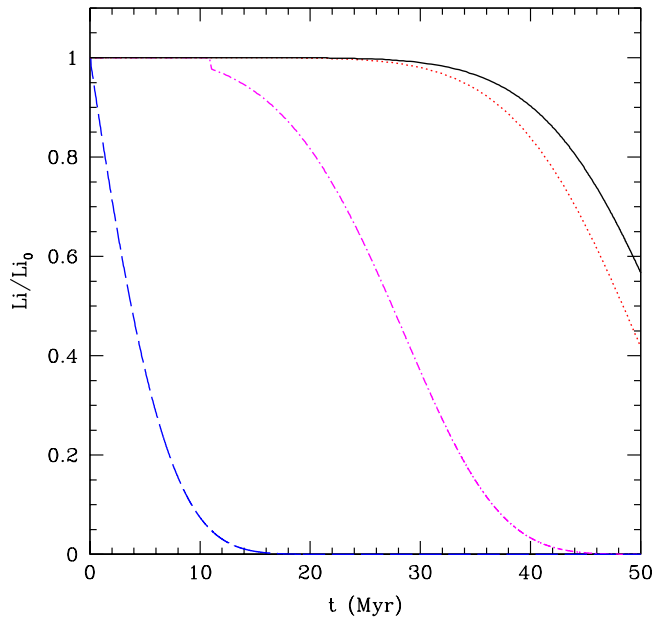,height=88mm,width=88mm}
\caption{Evolution of  the lithium abundance (divided by the initial Li abundance) as a function of time for the same models producing a 0.1 $\msol$ low-mass star as in Fig. \ref{mrtc_mf01}.}
\label{li_mf01}
\end{figure}

The significantly higher central temperatures induced by episodic accretion affect the timescale of lithium depletion, leading to faster
lithium depletion compared with a non accreting object of the same age,
as illustrated in Fig. \ref{li_mf01} for sequences producing 0.1 $\msol$ stars. Initial $\minit = 1 \mjup$ protostars experiencing burst accretion rates $\mdotburst = 5 \, 10^{-4} \msolyr$
produce a 0.1 $\msol$ object that entirely depletes its lithium
content within about 10 Myr, whereas for the non accreting counterpart, complete lithium depletion takes more than 50 Myr. As clearly illustrated in
Fig.  \ref{li_mf01}, depending on
the initial mass and the burst accretion rate, different episodic acccretion histories can 
produce objects with the same mass, say 0.1 $\msol$, at a same age, say $\simle$ 50 Myr, exhibiting {\it different levels of lithium depletion}. 
Similar effects are found for episodic accretion sequences producing final objects within the entire characteristic domain of fully convective stars, M $\simle \, 0.35 \msol$.

%In some cases, a small radiative core can develop at early stages for models produced by episodic accretion in this final mass regime when $\tc$ exceeds  $\sim 3 \, 10^6$ K. It however vanishes rapidly for models with final masses M $\simle 0.35 \msol$ as they  contracts toward the main sequence, given their large central density. For example, the 0.1 $\msol$ model produced by burst accretion starting with $\minit = 1 \mjup$ and with $\mdotburst = 5 \, 10^{-4} \msolyr$ (blue long-dashed curve in Figs. \ref{mrtc_mf01}) reaches $\tc \sim 3 \, 10^6$ K for a central density $\rho_{\rm c} \sim 70$ g cm$^{-3}$. In comparison, a 1 $\msol$ model produced by burst accretion starting with $\minit = 10 \mjup$ and with $\mdotburst = 5 \, 10^{-4} \msolyr$ (blue long-dashed curve in Figs. \ref{mrtc_mf1})), reaches this central temperature with a central density $\rho_{\rm c} \sim 1$ g cm$^{-3}$ (CHECK). Such density is low enough to allow the formation of a radiative core, given the sensitivity of the "metal bump" opacities at $T \sim 3 \, 10^6$ K to density (see Fig. 2a of Rogers \& Iglesias 1992). 
 
\subsection{The case of stars that develop a radiative core}

\begin{figure}
\psfig{file=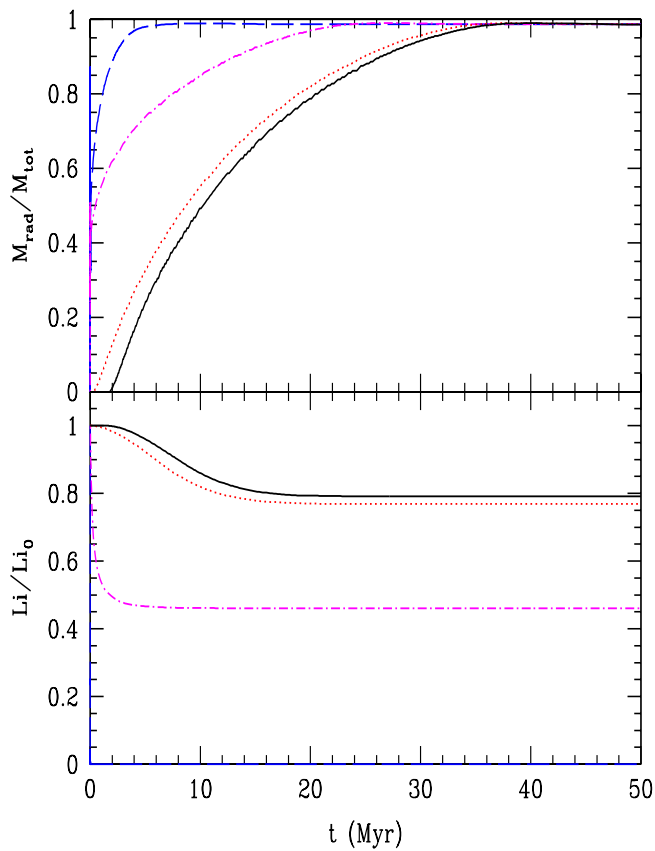,height=120mm,width=88mm}
\caption{Evolution of the mass of the radiative core divided by the total mass (upper panel)   and of the surface lithium abundance divided by the initial Li abundance (lower panel) as a function of time for models
producing a 1 $\msol$ star. The curves refer to the same models  as in Fig. \ref{mrtc_mf1}. Note that
Li is entirely depleted in the model with $\minit$ = 10 $\mjup$ (long-dashed blue line) in less than 1 Myr.
}
\label{li_mf1}
\end{figure}

If the central temperature exceeds $\sim$ 2-3 10$^6$ K, a radiative core develops,
because of the opacitiy decrease after the last ``$\kappa$-bump" due to metals (C, O, Ne and Fe, see Rogers \& Iglesias 1992), as explained in Chabrier \& Baraffe (1997, see their \S 3.2 and their Fig. 9). The exact temperature at which this occurs depends on the density, given the sensitivity of the opacities to density in this
temperature range (see Fig. 2a of Rogers \& Iglesias 1992). The higher the density, the higher the temperature required for the radiative core to develop.  Since this range of temperatures is also 
characteristic of the temperature required for Li nuclear fusion, the central 
temperature and density at which the radiative core develops will determine the temperature at the bottom of
the convective envelope and thus the level of
Li depletion in the convective envelope. For sequences producing  1 $\msol$ stars, Fig. \ref{li_mf1}
shows that accretion history has a strong impact on (i) the age for the onset of a radiative core, (ii) the mass of the convective envelope
at ages $\simle$ 30 Myr, and (iii) the Li abundance in the convective envelope. The higher central temperature reached by sequences with strong burst accretion rates (see Fig. \ref{mrtc_mf1})
results in
a radiative core that develops earlier in time. For the sequence starting from $\minit$ = 10 $\mjup$, its more compact structure yields significantly higher temperatures at the bottom of the convective envelope, with a maximum of $\sim$ 7 10$^6$ K, resulting in complete Li depletion in
the convective envelope at ages $<$ 1 Myr. In comparison, for the accreting sequence starting with $\minit$ = 0.1 $\msol$, the maximum temperature reached at the bottom of the convective envelope is $\sim$ 3.5 10$^6$ K, which is comparable to the value found in the non accreting sequence (see Fig. \ref{li_mf1}).
These differences in temperature are crucial for Li depletion, yielding very different final lithium abundances in the convective envelope, as illustrated in Fig. \ref{li_mf1}.
 
The main results of this section can be summarized as follows.  Depending on the initial seed mass and the burst rates, different episodic accretion histories can produce
1 $\msol$ star models with different sizes of the convective envelope at the same age, for ages $\simle$ 30 Myr.
The models can have different surface Li abundances {\it even after the models have converged
toward the same structure}, i.e. after $\sim$ 30 Myr in the particular case portrayed in Fig. \ref{li_mf1}.  
Similar effects are found for models with episodic accretion producing partly convective objects in the mass range 0.35 -1 $\msol$. 

\section{\label{discussion} Discussion}

\subsection{Summary of the results}

We have shown in the present work that an early protostar/BD
accretion history based on
episodes of short, intense bursts of accretion with typical accretion rates $\mdotburst = 10^{-4} \, - \, 5 \, 10^{-4} \msolyr$, as obtained in 2D hydrodynamical simulations of gravitationally unstable accretion disks (Vorobyov \& Basu 2005), can affect the internal structure of low-mass objects even after several Myr, up to a few times 10 Myr. The main results obtained in the present analysis can be summarized as follows.
\begin{itemize}
\item{} 
(i) Episodic accretion produces objects with smaller radius and higher central temperature compared
to the non accreting counterpart with the same mass at the same age.
\item{}  (ii) Higher central temperatures at a given age and mass can significantly enhance lithium depletion in fully convective low-mass stars.
\item{} (iii) As a consequence of the hotter structure, a radiative core develops earlier in accreting objects with final masses $M \simgr 0.35 \msol$. Episodic accretion can thus lead to final objects with smaller convective envelopes than predicted by standard non accreting models, at ages up to a few times 10 Myr. 
\item{} (iv) The more compact  and hotter structure of accreting models with $M \simgr 0.35 \msol$ increases the maximum temperature reached at the bottom of the convective envelope, compared to
 the non accreting counterpart, increasing the level of lithium depletion in the convective envelope.
\end{itemize}
We stress that {\it the magnitude of all these effects strongly depends
 on the initial mass, the acretion burst rate, and the total number of bursts},
 but barely depends on the phases of quiescent accretion, as long as $\dot{M}_{\rm quiet}\ll 10^{-6} \msolyr$.
 Within the range of parameters mentioned in \S \ref{episodic}, episodic accretion can produce final low-mass star models that depart at various levels from standard non accreting models in terms of radius, luminosity  (and thus position in the HRD, see BCG09), lithium abundance, and convective envelope mass, at a given similar age. Conversely, final objects issued from episodic accretion
display observational signatures similar to non-accreting objects of same mass, but {\it at a younger age}, as shown in BCG09, the age difference again depending  on the various accretion histories.

\subsection{Comparison to observations: anomalous lithium depletion in young clusters}

\begin{figure*}
\psfig{file=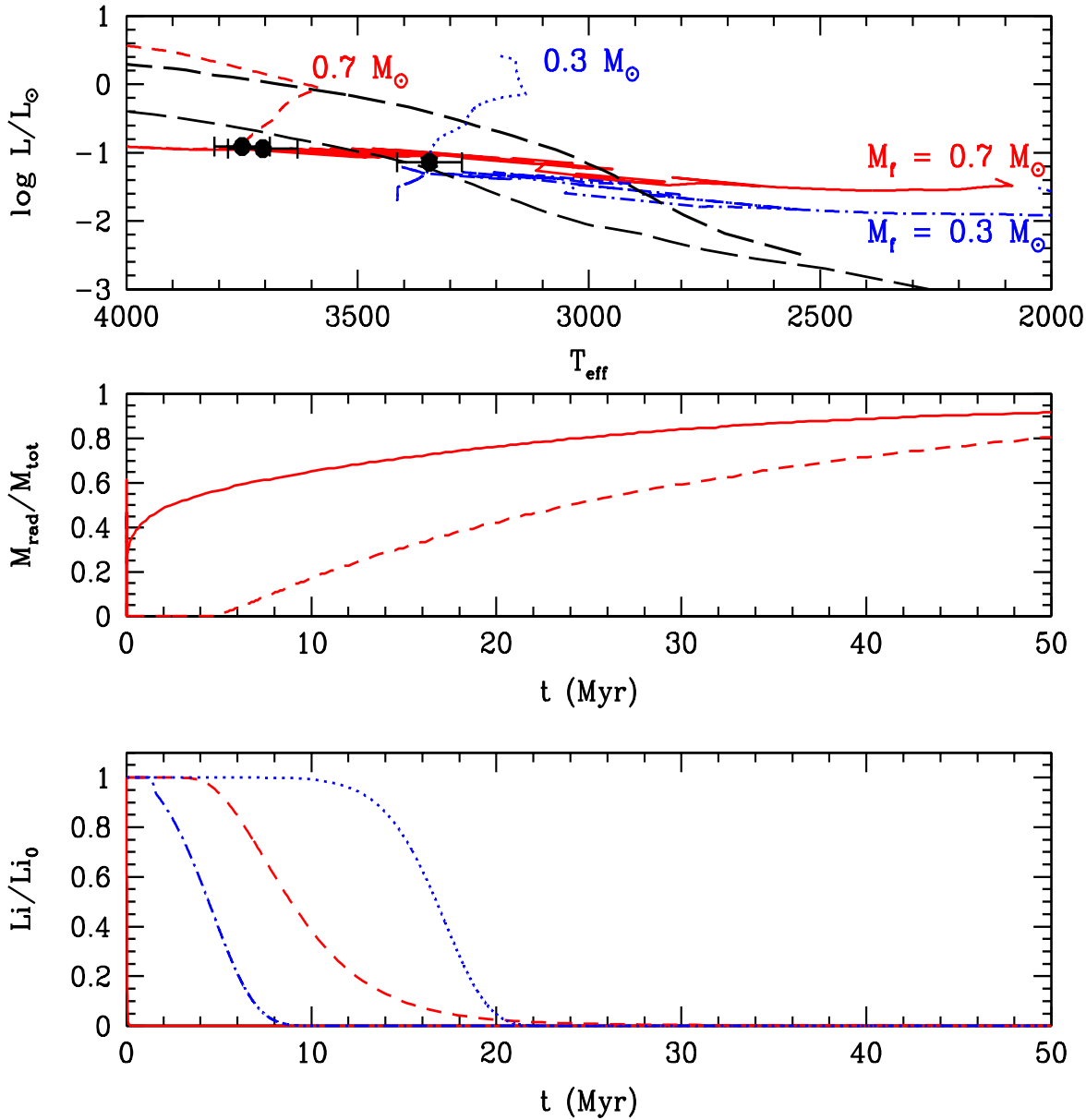,height=180mm,width=160mm}
\caption{Comparison of accreting models with observations of Sacco et al. (2007). Upper panel: HR diagram. The two long-dashed (black) curves are the 1 Myr and 10 Myr  isochrones of Baraffe et al. (1998)
for non accreting models.  The black points are the objects of Sacco et al. (2007). Middle panel: radiative core mass versus time for models producing a 0.7 $\msol$ star. Lower panel: surface Li abundance versus time.
The solid red curve is indistinguishable from the X-axis (i.e. $[Li]=0$): accreting sequence producing an object with final mass $M_{\rm f}$=0.7 $\msol$  with 
$\minit$ = 20 $\mjup$, $\nburst =$14; Dash-dotted blue: accreting sequence producing an object with final mass $M_{\rm f}$=0.3 $\msol$ with $\minit$ = 10 $\mjup$, $\nburst =$6. In both cases, $\mdotburst = 5 \, 10^{-4} \msolyr$, $\tburst$=100 yr and $\tquiet$=1000 yr; Dashed red: non accreting 0.7 $\msol$ model;  Dotted blue: non accreting 0.3 $\msol$ model. 
}
\label{hrd_sacco}
\end{figure*}

Several observational studies report discrepancies between ages inferred from evolutionary isochrones, on one hand,  and from lithium depletion, on the other hand, for M to G type stars of various ages (e.g Yee \& Jensen 2010). Observations also show lithium scatter in clusters and associations (e.g  da Silva et al. 2009; King  et al. 2010) and
young cluster members exhibiting severe, unexpected Li depletion for the inferred age of the cluster (e.g Kenyon et al. 2005; Sacco et al. 2007). The present calculations provide a consistent solution to
 these puzzles,  in particular providing an explanation
for the anomalous lithium depletion observed in some members of clusters that are a few Myr old (Kenyon et al. 2005; Sacco et al. 2007; Sacco et al. 2008; da Silva et al. 2009 ). 
The explanation suggested for these ``interlopers", found for instance in $\sigma$ Ori, 
$\lambda$ Ori, and the ONC, is to invoke a significant age spread within these clusters. However, at least in some cases (see below), the required age spread is significant, up to 10-20 Myr, a rather unlikely possibility in such young clusters. The present calculations, based on early evolution sequences involving accretion outbursts, provide a more plausible explanation, as illustrated in Fig. \ref{hrd_sacco} for the three weird objects
discovered by Sacco et al. (2007) in $\sigma$ Orionis. These objects show severe Li depletion, inconsistent with the age of the cluster (about 5 Myr), but have radial velocities consistent with cluster membership. They are displayed in the upper panel of Fig. \ref{hrd_sacco}. Evolutionary
sequences undergoing burst accretion with rates 
$\mdotburst \sim 5 \, 10^{-4} \msolyr$ and reaching final masses 0.3  $\msol$  and 0.7 $\msol$, respectively, reach the observed locations in the HRD at an age of $\sim$ 5 Myr,
consistent with the age of $\sigma$-Ori, whereas
the same locations correspond to 10-15 Myr for the non accreting sequences of same mass, as shown in the figure. For the  accreting sequence producing a 0.7 $\msol$ object, the middle panel of Fig. \ref{hrd_sacco} shows that a radiative core develops much earlier compared to the corresponding non accreting sequence. The lower panel illustrates the much faster Li depletion for the accreting sequences, as discussed in \S \ref{section_lithium}, with complete destruction occurring within less than 1 Myr for the sequence producing a 0.7 $\msol$ object. The possibility
for the accretion burst models to explain consistently both the position in the HRD
and the observed severe lithium depletion for these objects {\it at the cluster age} without particular fine tuning of the accretion parameters provides a strong support for this scenario. Although the high level of Li depletion in the three objects of
Sacco et al. (2007) has recently been questioned by Caballero (2010), similar anomalous Li-depletions  have been reported in other clusters by other groups (see references above), which can also be explained by episodic accretion. Our results thus reinforce the word of caution of  da Silva et al. (2010) 
regarding the elimination of stars as cluster members based only on Li abundances. The present analysis demonstrates that lithium is {\it not} a reliable age indicator, as its fate strongly depends on the past accretion history experienced by the star!

The same episodic accretion calculations can explain the unexpected degree of lithium depletion of stars with mass $>$ 0.5 $\msol$ in the young cluster IC 4665 (Jeffries et al. 2009), or the spread in Li abundance in cool dwarfs of about a solar mass reported
in the Pleiades (King et al. 2010), although other scenarios,  based either on the star rotational history (Bouvier 2008) or stellar activity (King et al. 2010), have been suggested. 
%We have not yet explored all issues and consequences of episodic accretion regarding lithium depletion and work is in progress to more systematically confront the present scenario  to these lithium problems observed in different clusters.

\subsection{Abundances in planet host stars}

Another interesting implication of our calculations and the inferred link between episodic accretion, lithium depletion, and the size of the convective envelope in young stars is the connection with the presence of close-in giant planets.  Zhu et al. (2010b) suggest that the
presence of mass clumps in some specific regions of circumstellar disks, which may lead eventually to non steady accretion when the disk becomes unstable (Vorobyov \& Basu 2005), could play an important role in the  process of planet formation. Following up along these lines,  we can speculate that the presence of giant planets, which require massive protoplanetary disks, more prone to gravitational instability, could be linked to the intensity of episodic accretion outburts. If this suggestion happens to be true, the present calculations provide an explanation for the claimed higher lithium depletion in planet host stars (Israelian et al. 2009; Sousa et al. 2010): since
these planet host stars will have undergone more intense accretion bursts, they will experience more intense Li depletion than similar planetless stars whose presumably less massive disks are less prone to violent accretion-burst episodes. 

Furthermore, as shown in this paper, episodic accretion can produce objects whith smaller convective envelopes than the ones obtained with standard non accreting models at ages $\simle$ 10 Myr (see Figs. \ref{li_mf1} and \ref{hrd_sacco}), the characteristic maximum lifetime of protoplanetary disks. Consequently, we can expect the signature of (both terrestrial and gaseous) planet formation to leave a greater imprint on stars that have undergone phases of intense accretion bursts. Although speculative, this suggestion provides a consistent explanation for the recently claimed peculiar chemical abundance of the Sun, with about 20\% depletion of refractory elements relative to volatile elements, compared with solar-type analogs harboring close-in giant planets (Melendez et al. 2009; Ramirez et al. 2009). According to these authors, this chemical peculiarity seems to be common among Sun-like stars with {\it no detected giant planets} and is said to reflect the signature of the formation of {\it terrestrial 
planets}, which trap the refractories, whereas the dust-depleted gas is accreted by the central star
(Ramirez et al. 2009; Melendez et al. 2009). For such a suggestion to be viable, and the signature of terrestrial planet formation to be imprinted in the Sun photospheric composition, the Sun's convective envelope, of mass $M_{\rm env}$, must have been much smaller during the lifetime of the accretion disk, i.e. during the first $\sim$ 10 Myr, than predicted by standard pre-main sequence models. 
%If not, the new stellar abundance of element $i$ in the convective envelope, $X''_i = X_i [1+(M_{acc}/M_{env})(X^\prime_i/X_i)]$, where $X^\prime_i$ and $X_i$ denote the (different) mass fractions of element $i$ in the accreting material and in the host star, respectively, and $M_{acc}$ the accreted mass,
%will barely differ from the star "standard" characteristic abundance. 
Indeed, let $X'_i$ and $X_i$ be the mass fractions of element $i$ in the accreting material and in the originally unpolluted star convective envelope, respectively. The new abundance $X''_i$ of such element in the convective envelope becomes
\begin{equation}
X''_i = {X_i \over 1+\alpha}(1+\alpha {X'_i \over X_i}),
\end{equation}
where $\alpha=M_{\rm acc}/M_{\rm env}$ is the ratio of the accreted mass to the star convective envelope mass.
The ratio of two elements, say $i \equiv$refractories and $j \equiv$volatiles, in the convective envelope after accretion of disk material becomes
\begin{equation}
{X''_i \over X''_j}= {X_i \over X_j} \, {1+\alpha {X'_i \over X_i} \over 1+\alpha {X'_j \over X_j}}.
\end{equation}
In the presence of terrestrial planets, we expect  $X'_i < X_i$ and $X'_j > X_j$, since most refractories remain trapped in the planet, and thus
 \begin{equation}
{X''_i \over X''_j} < {X_i \over X_j}.
\label{inequ}
\end{equation}
The larger $\alpha$, hence the smaller the convective envelope,  the larger the inequality Eq.(\ref{inequ}), hence the larger the indirect signature of planet formation in the star's convective envelope.

To obtain such a result, 
the aforementioned authors refer to the results of Wuchterl \& Klessen (2001), who explored the formation and early evolution of a 1 $\msol$ star, and found that the early Sun
was never fully convective. These calculations, however, are incorrect. First of all, as discussed in Baraffe et al. (2003),
the calculations of
Wuchterl \& Klessen (2001) yield an evolutionary track for a 1 $\msol$ star that is much too hot,
by $\sim 700-800$ K in $\te$ at a given $L$,
compared to observed young binary systems with (dynamically determined) masses around a solar mass. Second of all, the unrealistic assumption of spherical accretion in the Wuchterl \& Klessen calculations leads to significantly hotter inner structures, which will favor the growth of a radiative core,  compared with more realistic 3D rotational collapse calculations (Chabrier et al. 2007, Fig. 3). This assumption is thus certainly responsible for the above-mentioned large 
overestimate of the effective temperature at young ages. 
On the other hand, the effect of
episodic accretion on the mass of the convective envelope illustrated in Fig. \ref{li_mf1} provides a more plausible explanation for this problem, and suggests that the early Sun could have undergone a phase of strong accreting bursts. 

Although interesting, however, this interpretation of the peculiar refractory to volatile  abundance ratio in the Sun and other solar twins without detected close-in giant planets as a consequence of terrestrial planet formation is challenged by the recent analysis of Chavero et al. (2010), based on
four CoRoT planet host stars. This work suggests that alteration of abundances of refractories with respect to volatiles in stellar atmospheres may simply result from
condensation processes in the accretion disk and from accretion of such altered gas onto the star. Independent of the source of this peculiar abundance pattern, the signature of such condensation/accretion processes will still leave a greater imprint  if
the star convective envelope is considerably smaller than usually expected for solar type stars at ages $< $10 Myr. Also,  the analysis of Chavero et al. (2010) reveals no obvious correlation between abundances and
condensation temperatures in the  CoRoT stars, indicating no sign of overabundance of volatiles relative to refractories in stars harboring close-in giant planets compared to the Sun,  contradicting the
observations of Melendez et al. (2009) and Ramirez et al. (2009). The  peculiar abundance ratio determinations in
the Sun and solar analogs without detected giant planets must thus be confirmed by further studies. 

In conclusion, anomalous abundance ratios of refractories to volatiles in stars may not necessarily reflect the signature of terrestrial planet formation but
may also result from condensation processes in the disk. A detectable  signature of these abundance anomalies, however, requires a smaller convective envelope during typical disk lifetimes than predicted by standard models. Strong accretion outburts do produce these favorable conditions.

%Since the smaller convective envelope is linked with the stronger Li depletion, our calculations then predict a direct correlation between the underabundance of refractory elements and the level of lithium depletion in LMS and solar-type stellar atmospheres: stars exhibiting unexpectedly large depletion of refractory elements should also show trends of higher Li depletion. 
%{\bf a marche pas forcement: les etoiles avec des close-in gisant planets not probablement eu des Žpisodes de bursts, et donc seront depletees en Li (zone convective plus petite), mais comme elles auront moins de chance d'avoir des plantes terrestres, elles ne seront pas depletees en rŽfractaires ?}.

\section{Conclusion}

We have shown here that non steady accretion can strongly affect the internal mechanical (radius) and thermal (temperature) structures, hence the observational signatures ($L,\te$) of low-mass stars (and brown dwarfs), possibly up to ages of a few tens of Myr. The magnitude of the effects strongly depends on
the forming object's initial core mass, the number of bursts, and the intensity of the burst accretion episodes. 
As already highlighted in BCG09, episodic accretion can
produce models of a given final mass that to various extents depart from standard non accreting models in terms of position in the HRD, Li abundance, and convective envelope mass. 
Since the mechanisms responsible for the accreting disk formation and possible outbursts depend on many environmental parameters (e.g. mass, temperature, angular momentum, magnetic field of the parent collapsing cloud, and thermal properties of the disk),
it is certainly possible to produce various populations of stars at a similar age in young clusters and associations that depart {\it at various levels}
from predictions of standard pre-main sequence models. 
Therefore, our scenario does not {\it systematically} predict a broad
spread among these stellar properties/signatures, because the level of departure entirely depends on the star or cluster properties and accretion histories. Further
observational effort is needed to explore these issues in more detail and to try to have better constraints on the processes of non steady accretion during the early stages of evolution.  
Among the multiple consequences of early episodic accretion, we emphasize
the limited (to say the least) reliability of Li depletion as an age or cluster membership indicator. We urge observers to re-analyze young cluster populations on the basis of the present results, since several genuine cluster members may have been eliminated because of their
unexpectedly low Li abundance. Any information about the fraction of such Li depleted objects, with membership unambiguously assessed by radial velocity, could provide  crucial insight into the occurrence
and the intensity of non steady accretion. 

We also suggest that intense bursts of episodic accretion may have some link to planet formation, {\it if} planet formation preferentially occurs in massive disks, which are more prone to instability.
%We also show that episodic accretion may have some impact on planet formation and on the imprint of such a process in the atmospheric element abundances of planet-hosting stars. 
We speculate that stars harboring planets, including our Sun, may have
experienced burst accretion episodes during their youth, which then led to a hotter thermal structure, thus to faster lithium depletion and a smaller outer convective zone than conventionally admitted. This provides
a natural explanation for the higher lithium depletion  observed in stars harboring at least giant planets. This also provides a plausible, although admittedly speculative, explanation for the claimed underabundance of refractory elements in the atmosphere of stars with no detected close-in giant planets, as a result of the formation of terrestrial planets, which trap large amounts of refractory elements.
 %For the same reason, we also predict that stars exhibiting anomalously large depletion of refractory elements should also be severely lithium depleted, and we speculate that these signatures are related to early episodes of episodic burst accretion and thus - if indeed planet formation occurs preferentially in the most massive disks - to the occurrence of planet formation.
In any event, the multiple consequences of episodic accretion that we characterize in the present work should stimulate deeper theoretical and observational investigations, in order to better understand the twilight zone characteristic of (proto)star early evolution.

\begin{acknowledgements}
We are grateful to M. Asplund, J. Bouvier,  L. Hartmann, A. Morbidelli for valuable discussions. I. B and G. C thank the Max-Planck Institute  for Astrophysics of Garching, where part of this work was completed, for their warm hospitality. This work was supported by the Constellation European network MRTN-CT-2006-035890, the French ANR "Magnetic Protostars and Planets" (MAPP) project, and the "Programme National de Physique Stellaire" (PNPS) of CNRS/INSU, France. The authors visited KITP, Santa Barbara, during completion of this work, and this research was  supported in part by the National Science Foundation under Grant No. PHY05-51164.
\end{acknowledgements}

\end{document}